\documentclass[12pt]{article}
\pdfoutput=1
\usepackage{graphicx}
\usepackage{amsmath,amsfonts,amssymb,stmaryrd,latexsym}
\usepackage{url,subcaption,comment}
\usepackage{hyperref}
\usepackage{xcolor}
\hypersetup{
    colorlinks=blue,       
    linkcolor=purple,   
    citecolor=blue,        
    filecolor=magenta,      
    urlcolor=black           
}
\usepackage{cite}
\usepackage{dcolumn}
\usepackage{bm}
\usepackage{booktabs ,ragged2e}
\usepackage{diagbox}
\newcommand{\be}{\begin{equation}}
\newcommand{\ee}{\end{equation}}
\newcommand{\bea}{\begin{eqnarray}}
\newcommand{\eea}{\end{eqnarray}}
\newcommand{\bear}{\begin{eqnarray}}
\newcommand{\eear}{\end{eqnarray}}
\newcommand{\ba}{\begin{array}}
\newcommand{\ea}{\end{array}}

\newcommand{\phase}{{\beta}}

\begin{document}

\baselineskip=18pt \pagestyle{plain} \setcounter{page}{1}

\vspace*{-1.cm}

\noindent \makebox[11.9cm][l]{\small \hspace*{-.2cm} }{\small Fermilab-Pub-25-0593-T}  \\  [-1mm]

\begin{center}

{\Large \bf  Multi-top signals of vectorlike quarks at the LHC
} \\ [1cm]

{\normalsize \bf Elias Bernreuther$^{\diamond}$ and Bogdan A. Dobrescu$^\star$ \\ [5mm]
{\small {\it $^\diamond$Department of Physics, University of California, San Diego, La Jolla, CA 92093, USA     }}\\ [2mm]
{\small {\it $^\star$Particle Theory Department, Fermilab, Batavia, IL 60510, USA     }}\\[14mm]
}

\center{December 15, 2025}

\end{center}

\vspace*{0.8cm}

\begin{abstract} \normalsize
We point out that events with 6 or more top quarks may be observed at the LHC if certain particles exist at the TeV scale. In a model where a vectorlike quark of charge 2/3 decays into a top quark and a pseudoscalar particle, which subsequently decays into a top-antitop pair,  the LHC production cross section for events with 6 top quarks may be above 10 fb. If the pseudoscalar is part of a complex scalar field, then longer cascade decays, involving the scalar partner, may lead to events with 8 or even 10 top quarks.  We show that for a region of parameter space the dominant LHC signal in this model is 8 top quarks ({\it i.e.},  four $t\bar t $ pairs). The ensuing signals would be spectacular, including many leptons and $b$ jets. A discovery in that case would allow several cross section measurements that may determine the masses of all three new particles.
\end{abstract}

\newpage

\vspace*{1cm}

\renewcommand{\contentsname}{\normalsize\large Contents}
{ \hypersetup{linktocpage} 
\tableofcontents
\hypersetup{linkcolor=red} 
}

\bigskip

\section{Introduction}
\label{sec:intro}

Production of a top quark and a top antiquark (referred to as top pair production)  has a large cross section at hadron colliders, due to the sizable QCD coupling of the quarks to gluons. The top quark was discovered 30 years ago \cite{CDF:1995wbb, D0:1995jca} using this $t\bar t$ production at the Tevatron. In the Run 2 of the LHC, with proton-proton collisions at a center-of-mass energy of 13 TeV, the $t\bar t$ production cross section has been measured to be about $8.3 \times 10^5$ fb \cite{ATLAS:2023gsl}, in agreement with the SM computations \cite{Czakon:2013goa}.

The production of two $t\bar t$ pairs, commonly referred to as $4t$ production, proceeds in the Standard Model (SM) through the same QCD coupling and has a much smaller cross section  of approximately 13 fb in Run 2 \cite{vanBeekveld:2022hty, Frederix:2017wme}. Nevertheless, both the ATLAS and CMS Collaborations have recently observed $4t$ production in Run 2 of the LHC \cite{ATLAS:2023ajo, CMS:2023ftu}, with measured cross sections consistent with the SM prediction. 

It is interesting to investigate whether events with more than four top quarks or antiquarks may be observed at the LHC. 
The SM predicts that events with three $t\bar t$ pairs (which we refer to as ``$6t$ production") are produced with a cross section of roughly $4 \times 10^{-4}$ fb, based on our leading-order computation with \textsc{MadGraph} \cite{Alwall:2014hca}. This implies that by the end of the High-Luminosity LHC runs, with an integrated luminosity of 3000 fb$^{-1}$, barely one $6t$ event will be  expected to be produced by SM processes. Given the various backgrounds, from QCD, $W$ + jets, $WW$ + jets, $t\bar t$ + jets, and other processes, observation of events with more than two $t\bar t$ pairs would require a cross section larger than the SM one by at least two orders of magnitude, and would thus be striking evidence of new physics.

Here we show that a simple extension of the SM may sufficiently enhance the production of events with $6t$, $8t$ ({\it i.e.}, four $t\bar t$ pairs),  and even $10t$ ({\it i.e.}, five $t\bar t$ pairs) so that they can be observed at the LHC.
We focus on a renormalizable model that includes a vectorlike quark ($\chi$) of electric charge 2/3 and a gauge-singlet complex scalar ($\phi$). Both these fields are well motivated by a variety of theories beyond the SM. 
For example, the vectorlike quark $\chi$ is a necessary ingredient in composite Higgs models, in which it is typically accompanied by at least one gauge-singlet complex scalar that is a quark-antiquark bound state \cite{Dobrescu:1999gv, Cheng:2013qwa, Cheng:2014dwa}. Another example is a theory of quark and lepton compositeness \cite{Dobrescu:2021fny, Assi:2022jwg}, in which $\chi$ arises as a three-preon bound state and $\phi$ is a six-preon bound state.

The vectorlike quark has mass mixing with the SM quarks of charge 2/3, with the effect expected to be the largest for the third generation. Thus, the mixing between $\chi$ and the third-generation up-type quark gives two mass eigenstates: the observed top quark, and a heavy quark $t'$. 
The complex scalar field $\phi$ contains two spin-0 particles: a pseudoscalar $a_t$, and a scalar $\varphi_t$. If kinematically allowed, the $t'$ quark may decay into one of these spin-0 particles and a top quark, while $a_t$ decays into $t \bar t$.

The pseudoscalar $a_t$ may be a pseudo-Nambu-Goldstone associated with an approximate global symmetry, and thus it is natural to expect that the masses of the two new spin-0 particles satisfy $m_\varphi > 2 m_a $. As a result, the main decay mode of the scalar particle  is $\varphi_t \to a_t \, a_t$. 
Therefore, pair production of the vectorlike quark is followed by  cascade decays such as $t' \to t \, a_t \to t (t \bar t) $,  or  
$t' \to t \, \varphi_t \to t \, a_t  \, a_t  \to t  (t \bar t)( t \bar t) $, resulting in 3, 4 or 5 $t \bar t$ pairs. It is remarkable that such a simple and well-motivated extension of the SM gives rise to long cascade decays and very high multiplicity final states, which the LHC can already begin to probe.

LHC signals with three $t \bar t $ pairs have been previously proposed \cite{Han:2018hcu,Bhardwaj:2022nko,Dermisek:2019vkc,Deandrea:2014raa,Bizot:2018tds,Dermisek:2020gbr}. It turns out that our $8t$ signal\footnote{A different $8t$ signal was considered in \cite{Deandrea:2014raa} based on coloron pair production \cite{Dobrescu:2007yp,Cacciapaglia:2024wdn}.} 
typically (for a large enough mass ratio $m_{t'} / M_\varphi $) has a larger total branching fraction than the $6t$ signal. Furthermore, all three types of signal processes analyzed here ($6t$, $8t$, and $10t$) give rise to comparable number of events, provided the mass ordering $ m_{t'} > M_\varphi + m_t > 2 M_a + m_t > 5 m_t$ does not involve near degeneracies. Hence, search strategies should be devised such that events due to the three signal processes can be statistically disentangled.

In Section~\ref{sec:model} we present the renormalizable model, and derive the branching fractions for the $t'$ quark and the spin-0 particles. We discuss the  LHC signals and possible search strategies in Section~\ref{sec:8top}. Our conclusions are summarized in Section~\ref{sec:conc}.

\bigskip

\section{Weak-singlet vectorlike quark plus a complex scalar}
\label{sec:model}
\setcounter{equation}{0}

We consider a model where there are two fields beyond the SM: a vectorlike fermion $\chi$ transforming as $(3,1, +2/3)$ under the  $\mathrm{SU} (3)_c\times \mathrm{SU}(2)_W \times \mathrm{U}(1)_Y $ gauge group, and a complex scalar $\phi$, which is a gauge singlet. 

\subsection{Mass mixing of heavy quarks}
\label{sec:mass}

Given the charge assignment for the two new fields, Yukawa interactions of the singlet scalar $\phi$  with the vectorlike quark $\chi$ and third generation SM quarks can be written as
\begin{equation}
	-  \phi \, \overline {\chi}_L  \left(  y_\chi \, e^{i\phase_\chi}  \,   \chi_R  + y_o \, e^{i \phase_o }  \, u_R^3 \right)  +  {\rm H. c.}     ~~,
\label{eq:Yukawas}
\end{equation}
where $u_R^3$ denotes the SM right-handed up-type quark of the third generation, in a basis where the SM Higgs has Yukawa couplings that are diagonal in quark flavor. Both the diagonal and the off-diagonal Yukawa couplings ($y_\chi $ and $y_o$ respectively) are positive, and the two complex phases ($\phase_\chi$ and $\phase_o$) are in the $(-\pi,\pi]$ interval.
Similar Yukawa interactions involving second and first generation right-handed up-type quarks may exist, but here we will assume for simplicity that their effects are negligible. This assumption is motivated by the well-known fact that the Yukawa couplings of the SM Higgs doublet to second and first generations are suppressed.

In addition, the gauge charges of $\chi$ allow for the explicit mass terms
\begin{equation}
	- \overline \chi_L \left( m_{\chi\chi} \, e^{i \alpha_\chi}  \,  \chi_R + m_{\chi 3}  \, e^{i \alpha_o } \, u_R^3 \right) +  {\rm H. c.} \; ,
\label{eq:mass-terms}
\end{equation}
where $m_{\chi\chi}$  and $m_{\chi 3}$ are positive mass parameters, while $\alpha_\chi$ and $\alpha_o$ are phases in the $(-\pi,\pi]$ interval. We again focus on the case where Lagrangian terms involving the second and first generation quarks are negligible.
If the complex scalar has a VEV, $\langle \phi \rangle = v_\phi > 0$, then the Yukawa terms in (\ref{eq:Yukawas}) combine with the mass terms in (\ref{eq:mass-terms}) to generate effective mass terms involving $\chi$. Their mass parameters may be chosen to be positive upon separate field redefinitions of $\chi_L$ and $\chi_R$, so that the effective mass terms take the form
\be
- m_\chi \, \overline \chi \,   \chi - m_o \left( \overline {\chi}_L \,  u_R^3  + \overline u_R^3   \,   {\chi}_L\right)  \; .
\label{eq:effective-terms}
\ee
The above effective mass parameters are related to the parameters in (\ref{eq:Yukawas}) and (\ref{eq:mass-terms})  by $m_\chi = m_{\chi\chi} \, e^{i \alpha_\chi}  + y_\chi \, v_\phi  \, e^{i\phase_\chi}  > 0$ and $m_o = m_{\chi 3} \, e^{i \alpha_o} + y_o \, v_\phi  \, e^{i\phase_o}  > 0$. 

As already mentioned, the SM up-type quark fields in the above Lagrangian terms are defined in a mass-diagonal basis, so that their  interactions with the SM Higgs doublet $H$ (taken here to have hypercharge $-1/2$) are given by 
\begin{equation}
	- y_{u_j} \,  \overline{q}_L^j \, H \,  u_R^j  + {\rm H.c.}\;  ,
\label{eq:Higgs}	
\end{equation}
where $q_L^j$ is the quark doublet of the $j$th generation ($j = 1, 2, 3$), and $y_{u_j} > 0$. Besides the terms involving SM quarks, there may exist similar Yukawa interactions of the SM Higgs in which the $u_R^j$ fields are replaced by $\chi_R$. As for (\ref{eq:Yukawas}) and (\ref{eq:mass-terms}), we assume that those interactions of $\chi$ have negligible coefficients in the case of $q_L^2$ and $q_L^1$. For the third generation, the term $ \overline{q}_L^3 \, H \,  \chi_R$ may generically have a large coefficient; however, this term can be eliminated by a global  $\mathrm{U}(2)$ transformation acting on the $u_R^3$ and $\chi_R$ fields. Thus, without loss of generality we set the coefficient of $ \overline{q}_L^3 \, H \,  \chi_R$ to zero by choosing the appropriate basis for the $u_R^3$ and $\chi_R$ fields that appear in (\ref{eq:Yukawas})-(\ref{eq:Higgs}).    

Replacing $\phi$ and $H$ by their VEVs, and also the third generation quark doublet $q_L^3$ by its components, $(u_L^3, d_L^3)$, in (\ref{eq:Higgs}), we find the following matrix form for the mass terms of $u^3$ and $\chi$:
\begin{equation} \hspace*{-0.4cm}
- \left( \overline{u}^3_L \, , \;  \overline  \chi_L  \right)   
\left( \! \ba{cc}   y_{u_3}  v_H   &    0 
\\  [4mm]  
m_o \;
& \;  m_\chi 
  \ea  \! \right)  
\left( \ba{c}   u^3_R  \\ [3mm]   \chi_R  \ea  \right)   + {\rm H.c.} 
\label{eq:vquark_massmatrix1}
\end{equation}
Here $v_H \approx 174$ GeV is the electroweak scale, $m_\chi$ and $m_o$ are the mass parameters  from (\ref{eq:effective-terms}), and $y_{u_3}  > 0$, so all entries of the above mass matrix are positive. A global $\mathrm{SU}(2)_L \times \mathrm{SU}(2)_R$ transformation,
\begin{equation}
\left(  \!  \ba{c}   u^3_{L}  \\  \chi_{L}   \ea   \!  \right)   = 
\left( \! \ba{cc}  c_{_{L}}   &   s_{_{L}}    \\ [2mm]    -s_{_{L}}    &   c_{_{L}}   \ea   \!  \right)  
\left(  \!  \ba{c}   t_{L}  \\  t^\prime_{L}    \ea   \!  \right)    \;\;\; ,   \;\;\; \;\;
\left(  \!  \ba{c}   u^3_{R}  \\  \chi_{R}   \ea   \!  \right)   = 
\left(  \!  \ba{cc}  c_{_{R}}   &   s_{_{R}}    \\ [2mm]    -s_{_{R}}    &   c_{_{R}}   \ea   \!  \right)  
\left(  \!  \ba{c}   t_{R}  \\  t^\prime_{R}    \ea   \!  \right)    ~~,   
\label{eq:transf}
\end{equation}
rotates the gauge eigenstates $u^3_{L,R}$, $\chi_{L,R}$ into the mass eigenstates $t_{L,R}$ and $t^\prime_{L,R}$. The notation used here is $c_{_{L,R}} \equiv \cos\theta_{_{L,R}} $ and $s_{_{L,R}}  \equiv \sin\theta_{_{L,R}} $, and it is sufficient to consider the angular range $\theta_{_{L}}, \theta_{_{R}} \in [0, \pi/2)$. 
Requiring that transformation (\ref{eq:transf}) diagonalizes the mass terms \eqref{eq:vquark_massmatrix1} leads to
a relation between $\theta_{\! _R}$ and $\theta_{\! _L}$,
\begin{equation}
	\tan\theta_{\! _R}  =   \frac{m_{t^\prime} }{m_t}    \,    \tan\theta_{\! _L}    ~~,
\label{eq:tanR}
\end{equation}
as well as three relations (see \cite{Bernreuther:2023uxh} for a direct method to derive them) 
between the Lagrangian parameters ($m_\chi $, $m_o$, $y_{u_3}$) and the physical parameters (the $m_t$, $m_{t^\prime}$ masses and  $s_{\! _L}$ mixing). One of them determines the Yukawa coupling of the third generation up-type quark to the Higgs doublet:
\be
	\label{eq:relations_y}
  	y_{u_3}    =   \frac{m_t  \, c_{\! _L} }{v_H \, c_{\! _R}}   ~~,
\ee
The remaining two relations determine the mass parameters from (\ref{eq:effective-terms}):
\bear
&&	m_\chi = m_{t^\prime}   \frac{c_{\! _R}}{c_{\! _L} }    ~~, 
	\nonumber \\ [-2mm]
		\label{eq:relations_mixingangles}
	\\ [-2mm]	
&&  	m_o = m_{t^\prime} \frac{s_{\! _R}}{c_{\! _L} } - m_t \frac{s_{\! _L} }{c_{\! _R}}   ~~.
\nonumber
\eear

Given that $m_{t^\prime} \gg m_t$, it is possible to derive a strict upper limit on the mixing  $s_{\! _L}$  based on unitarity considerations \cite{Dobrescu:2009vz}. To obtain that limit, note first that the expression for $\theta_R$ from (\ref{eq:tanR}) implies that we can rewrite  (\ref{eq:relations_y}) as
\be
y_{u_3}    =   \frac{m_t }{v_H }  \left[  s_{\! _L}^2  \! \left(  \frac{m_{t^\prime}^2}{m_t^2} - 1 \right) + 1 \right]^{\! 1/2} ~~.
\label{eq:yu3sL}
\ee
A Yukawa coupling like $y_{u_3}$ increases logarithmically with the energy scale, and must be perturbative at least up to a scale where the SM Higgs boson arises from some underlying theory. 
Thus, we need to impose an upper limit, labeled $y_{\rm max}$, on the Yukawa coupling $y_{u_3}$ at the $m_t$ scale.
A rough estimate of $y_{\rm max}$ can be obtained based on the one-loop RGEs (see, {\it e.g.}, \cite{Hill:1980sq}) for the Yukawa coupling and the QCD coupling $g_s$,
\bear
\label{eq:RGEs}
&& \frac{d y_{u_3} }{d \ln \mu} = \frac{y_{u_3}}{16 \pi^2}  \left( \frac{9}{2} y_{u_3}^2 - 8 g_s^2 \right)   ~~,
\nonumber \\ [-2mm]
\\ [-2mm]
&& \frac{d g_s }{d \ln \mu} =  \frac{g_s^3} {16 \pi^2}  \left( 11 - \frac{2}{3} n_f \right)  ~~,
\nonumber 
\eear
where $\mu$ is the energy scale, and the number of quark flavors is $n_f = 6$ for $m_t < \mu <m_{t'}$, and $n_f = 7$ for $\mu > m_{t'}$. 
Let us denote by $\Lambda$ the UV scale where the Yukawa interaction $\overline{u}^3_L \, H \, u^3_R$ is blowing up ({\it i.e.}, it is no longer well defined, for example, due to Higgs compositeness \cite{Dobrescu:1997nm,Chivukula:1998wd}).
Solving numerically the set of differential equations (\ref{eq:RGEs}) with the boundary condition $y_{u_3}(\Lambda)  < 4\pi$ gives
\be
y_{u_3} < y_{\rm max} \approx 2.3  ~~
\ee
for $\Lambda = 10$ TeV.  When $\Lambda$ is varied in the $5-20$ TeV range, $y_{\rm max}$ decreases from 2.5 to 2.2, and it is relatively insensitive to $m_{t'}$. Note that although  $y_{u_3}$ is the SM Yukawa coupling of the top quark, values $y_{u_3} > 2$ in the model discussed here are still consistent with the measured top quark due to a seesaw mechanism in the $u^3-\chi$ system \cite{Dobrescu:1997nm}.

\begin{figure}[t!]
	\begin{center}
		\includegraphics[width=0.57\textwidth]{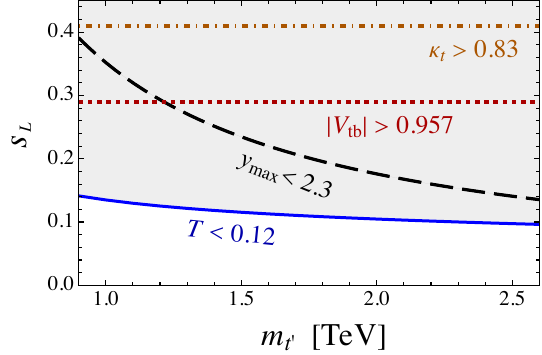}
		\caption{\label{fig:sL} Upper limits on the $s_L$ mixing due to the unitarity limit (\ref{eq:upper}) with  $y_{\rm max} = 2.3$ (dashed black line), the electroweak parameter $T < 0.12$ (solid blue line), the measured $|V_{tb}|$ element (dotted red line), and the measured Higgs-top coupling with a coupling modifier $\kappa_t > 0.83$ (dot-dashed orange line). The ruled out region lies above the solid blue line and it is shaded.}
	\end{center}
\end{figure}

The upper limit on $y_{u_3}$ in conjunction with (\ref{eq:yu3sL}) implies 
\be
 s_{\! _L}  <   \frac{m_t}{m_{t^\prime}} \left( y_{\rm max}^2 \,  \frac{v_H^2 } {m_t^2} - 1 \right)^{\! 1/2} ~~.
\ee
Here we took advantage of the lower mass limit, discussed in Section \ref{sec:8top}, $m_{t^\prime} \gtrsim 1$ TeV, by keeping only the leading term in $(m_t/m_{t^\prime})^2 < 3.0 \times 10^{-2}$. 
Due to the SM peculiarity that the electroweak scale is very close to the top quark mass ($v_H \approx m_t$), we can write the limit on mixing as
\be
 s_{\! _L}  <   0.17 \sqrt{ y_{\rm max}^2  - 1 } \; \frac{1 \; \rm TeV}{m_{t^\prime}}  ~~.
 \label{eq:upper}
\ee
This theoretical upper limit on $s_L$ is plotted (together with experimental limits discussed below) as a function of $m_{t^\prime}$ in Figure~\ref{fig:sL}.

Mass mixing between the third-generation up-type SM quarks and the vectorlike quark induces an off-diagonal coupling of the SM Higgs boson,  $h^0$, given by
\begin{equation}
	-  \frac{s_{\! _L} c_{\! _L} m_{t'} }{\sqrt{2} \, v_H } \; h^0 \,    \overline{t}_L	 \,  t^\prime_R + {\rm H.c.}   ~~,
\label{eq:Higgs-offdiagonal}	
\end{equation}
where we again eliminated $\theta_R$ and the Yukawa coupling using (\ref{eq:tanR}) and (\ref{eq:relations_y}), respectively. 
There is also an $h^0 \, \overline{t}_R t^\prime_L$ coupling, but we do not show it here because it is suppressed \cite{Dobrescu:2009vz} by an extra factor of  $m_t/ m_{t'}$.  Likewise, the diagonal coupling $h^0 \, \overline{t'} t'$ is too suppressed, by an $s_{\! _L}^2$ factor, to be phenomenologically relevant.
Compared to the SM, the diagonal coupling of the top quark to the Higgs boson is modified by an overall factor of $c_{\! _L}^2$, due to the mixing with $t'$:
\be
-  \frac{c_{\! _L}^2 \, m_{t} }{\sqrt{2} \, v_H } \; h^0 \,    \overline{t} \,  t  ~~.
\ee
The current measured value \cite{ATLAS:2025qxq} of the Higgs boson coupling modifier to the top quark is $\kappa_t = 0.99^{+0.10}_{-0.08}$ (the CMS 95\% CL lower limit is almost the same, see Table 11 of \cite{CMS:2025jwz}), so the 95\% CL lower limit is $c_{\! _L}^2 > 0.83$, which implies $s_{\! _L} < 0.41$.

Furthermore,  the mixing induces a $W$-$t^\prime$ interaction with the $b$ quark, as well as a modified (compared to the SM) $W$-$t$ interaction:
\begin{equation}
	\frac{g}{\sqrt{2}} \, V_{tb} \,  W^-_\mu \,  \left(c_{\! _L} \overline{t} + s_{\! _L} \overline{t^\prime} \right)  \gamma^\mu  P_L \, b 
	+ {\rm H.c.} 
\label{eq:W-offdiagonal}		
\end{equation}
Since the SM coupling of the $W$ boson to top and bottom quarks is slightly suppressed by a factor of $c_{\! _L}$, it follows that measurements of $V_{tb}$ impose an upper limit on $s_{\! _L}$. The average \cite{ParticleDataGroup:2024cfk} of LHC and Tevatron measurements of $V_{tb}$ from single-top production is given by $|V_{tb}| = 1.010\pm 0.027$. This yields a 95\% CL lower bound of $|V_{tb}| > 0.957$, translating to $s_{\! _L} < 0.29$. 

The off-diagonal interaction of the heavy quarks with the $Z$ boson reads
\begin{equation}
	\frac{s_{\! _L} c_{\! _L}  \, g}{2\cos\theta_W} \,   Z_\mu \, 
	\overline t_L   \gamma^\mu  t_L^\prime  + {\rm H.c.}    ~~,
\label{eq:Z-offdiagonal}	
\end{equation}
while the diagonal interactions are
\begin{equation}
	\frac{g}{\cos\theta_W}   Z_\mu \,   \left[ \overline t   \gamma^\mu  
	\left(\frac{c_{\! _L}^2}{2} P_L - \frac{2}{3} \sin^2\!\theta_W \right) t
	+ \overline t^\prime    \gamma^\mu  
	\left(\frac{s_{\! _L}^2}{2} P_L - \frac{2}{3} \sin^2\!\theta_W \right) t^\prime  \right]    ~~.
\end{equation}
This deviation from the SM $Z$ coupling, as well as the new couplings to $t'$, contributes to the electroweak $T$ parameter \cite{Chivukula:1998wd, Dobrescu:2009vz}:
\be
T = \frac{3 s_{\! _L}^2 \, m_t^2}{16\pi \, M_W^2 \sin^2\!\theta_W} \left[ \left( s_{\! _L}  \frac{m_{t'}}{m_t} \right)^{\! 2} 
+ 4 c_{\! _L}^2 \ln\! \left(\frac{m_{t'}}{m_t}\right) - 2 + s_{\! _L}^2  \right]~~,
\ee 
where we dropped corrections of order $(m_t/m_{t'})^2$. The fit to the electroweak data gives $T < 0.12$ at the 95\% CL \cite{ParticleDataGroup:2024cfk}, which compared to the above expression on $T$ leads to an upper limit on $s_{\! _L}$ that is monotonically  relaxing for larger $m_{t^\prime}$:
at $m_{t^\prime} = 1$ TeV the limit is $s_{\! _L}  <  0.11$; at $m_{t^\prime} = 2$ TeV the limit becomes $s_{\! _L}  <  0.14$.
This upper limit on $s_{\! _L}$, due to the fit to the electroweak observables, is shown as the solid blue curve in Figure~\ref{fig:sL}. 

The upper  limits on $s_{\! _L}$ derived above from the measured CKM element $V_{tb}$ and the measured top-Higgs coupling are also shown in Figure~\ref{fig:sL}, as the dotted red line and the dot-dashed orange line, respectively. Based on the most stringent limit, which is the one from electroweak data, it is justified that in the next Section we set $c_{\! _L} =1$ and also keep only the leading term in
$s_{\! _L}^2$. Note though that the upper limit on $s_{\! _L}$ is of the order of $m_t/m_{t^\prime}$, so we will not expand in  $s_{\! _L} m_{t^\prime}/m_t$ except when we explicitly assume that $s_{\! _L}$ is much smaller than the current limit. Furthermore, we will not expand in $s_{\! _L}$ in the cases where different powers $s_{\! _L}$ have coefficients given by the unknown coupling ratio $y_o/y_\chi$.

Due to its mixing with the $t$ quark,  the heavier $t^\prime$ quark can decay into SM particles. To leading order in $s_{\! _L}^2$ and in $m_t^2/m_{t^\prime}^2$, the partial widths of the decays induced by the off-diagonal couplings (\ref{eq:W-offdiagonal}),  (\ref{eq:Z-offdiagonal}) and (\ref{eq:Higgs-offdiagonal}) are given by 
\begin{equation}
	\Gamma \left( t^\prime \to b  W^+ \right) = 2 \Gamma \left(t^\prime \to t  Z^0 \right) = 2 \Gamma \left(t^\prime \to t  h^0  \right) 
	= \frac{s_{\! _L}^2 m_{t^\prime}^3}{32 \pi v_H^2} 	 ~~.
	 \label{eq:standard}
\end{equation}
The ratios of these widths obey the Goldstone equivalence theorem \cite{Han:2003wu}.
As we will see in the next Section, there are additional decay modes of the $t^\prime$ quark in the presence of the complex scalar $\phi$.

\subsection{Interactions and decays of the new spinless particles}  
\label{sec:mass}

The complex scalar $\phi$ includes two particles, a scalar $\varphi_t$ and a pseudoscalar $a_t$, and can be linearly parametrized as
\begin{equation}
	\phi =  v_\phi + \frac{1}{\sqrt{2} } \, \left( \varphi_t + i a_t\right)  
	~~.
\end{equation}
The terminology `pseudoscalar' refers to the fact that $a_t$ is odd under a parity transformation when the interactions of $\phi$ are parity-invariant. If they are not parity-invariant but they are invariant under CP transformations (see {\it e.g}., \cite{Dobrescu:1999gv}), then it is more appropriate to refer to $a_t$ as a CP-odd spin-0 particle, and to $\varphi_t$ as a CP-even scalar. Another term for particles like $a_t$, frequently used in recent years (see {\it e.g}., \cite{Blasi:2023hvb}), 
is axion-like particle (ALP). While ALP and pseudoscalar are almost synonymous, we prefer to use ALP only for light (GeV scale or below) pseudoscalars. Here we focus mainly on $a_t$ above the $t\bar t$ threshold (its mass satisfies $M_a \gtrsim 345$ GeV), so we refer to it as a pseudoscalar. We assume that the $\phi$ potential is CP-conserving, so that there is no tree-level mass mixing between $a_t$ and $\varphi_t$. 
We will consider interactions of $\phi$ with the heavy quarks that include complex phases, which induce $a_t-\varphi_t$ mixing at one loop; it is a good approximation to ignore this small mixing here.

Using the relation   (\ref{eq:transf}) between gauge and mass eigenstates, (\ref{eq:Yukawas}) implies
the following off-diagonal interactions of the scalar $\varphi_t$ and the pseudoscalar $a_t$ with the  top and $t^\prime$ quarks:
\begin{equation}
\label{eq:scalar_t_tprime_coupling}
- c_{\! _R} {\cal Y}   \, \frac{ \varphi_t + i a_t}{\sqrt{2} } \;   \overline{t}^\prime   P_R  \,  t + {\rm H.c.}  ~~,
\end{equation}
where we ignored terms suppressed by additional factors of $s_{\! _L}$, and we defined an effective Yukawa coupling,
\be
{\cal Y} \equiv  y_o   \, e^{i \phase_o }   -   y_\chi \, s_{\! _L} \, \frac{m_{t'}}{m_t} e^{i  \phase_\chi } 
~~,
\label{eq:effYuk}
\ee
which may be complex.

The corresponding diagonal interactions read
\begin{equation}
	s_{\! _L} c_{\! _R} {\cal Y} \,  \frac{ \varphi_t + i a_t}{\sqrt{2} } \;   \overline{t} P_R \, t + \mathrm{H.c.} \; 
\label{eq:scalar_t_coupling}
\end{equation}
for the top quark, and 
\begin{equation}
	- c_{\! _R} {\cal Y}^{\,\prime}  \, \frac{ \varphi_t + i a_t}{\sqrt{2} } \;  \overline{t^\prime}  P_R \, t^\prime + \mathrm{H.c.} \; ,
\label{eq:scalar_tprime_coupling}
\end{equation}
for the $t'$ quark, where we defined a second (also complex) effective Yukawa coupling:
\be
{\cal Y}^{\,\prime}  \equiv  y_\chi \, e^{i \phase_\chi}  +  y_o \, s_{\! _L} \, \frac{m_{t'}}{m_t} \, e^{i \phase_o} 
~~.
\label{eq:effYukprime}
\ee
While it is convenient to keep overall factors of $c_{\! _R}$ in the above Lagrangian terms, it is useful to 
mention that (\ref{eq:tanR}) implies
\be
c_{\! _R} = \left( 1 + s_{\! _L}^2 \, \frac{m_{t'}^2 }{m_t^2}  \right)^{\! -1/2}  ~~,
\ee
where we kept only the leading term in $s_{\! _L}^2$. As a result, there is a lower limit on $c_{\! _R}$, which decreases monotonically with $m_{t'}$ and is obtained when $s_{\! _L}$ is at its upper limit given by the solid blue curve in Figure~\ref{fig:sL}. For $m_{t'} =1 $ TeV the limit is $c_{\! _R} > 0.79$, while for $m_{t'}  =2 $ TeV it becomes $c_{\! _R} > 0.64$.

In addition to the `standard' decay modes (\ref{eq:standard}), $t^\prime$ can decay into a top quark and the pseudoscalar $a_t$ due to the off-diagonal coupling  (\ref{eq:scalar_t_tprime_coupling}), with a partial width
\begin{equation}
\label{eq:tprime_to_tat_width}
	\Gamma \!\left( t^\prime \to t  a_t  \right) =\frac{m_{t^\prime}}{64 \pi}  \, c_{\! _R}^2  
	\left| {\cal Y}  \right|^2  
	\left( 1 -  \frac{M_{a}^2}{m_{t^\prime}^2} \right)^{\! 2} ~~,
\end{equation}
where $M_a$ denotes the mass of $a_t$, and we neglected terms suppressed by  $m_t/m_{t'}$ (including $m_t$ corrections to the phase space). The same expression applies to the width of the $t^\prime$ decay into $t$ and the scalar $\varphi_t$,  
$\Gamma \!\left( t^\prime \to t  \varphi_t  \right)$, except for a $M_a \to M_\varphi$ replacement in the phase-space factor.
Thus, $\Gamma \!\left( t^\prime \to t  \varphi_t  \right) < \Gamma \!\left( t^\prime \to t  a_t  \right)$ for $M_\varphi > M_a$.

In the limit of $s_{\! _L}^2 \ll 1$ and $m_{t^\prime}^2 \gg m_t^2$ (which also implies $m_{t^\prime}^2 \gg M_h^2, M_Z^2,\,M_W^2$),  the $t'$ width for the decay involving the new pseudoscalar and the sum of the $t'$ widths into standard modes have a ratio
\be \hspace*{-0.4cm}
R_t \equiv  \frac{\Gamma \!\left( t^\prime \! \to t  a_t  \right) }{\Gamma \!\left( t^\prime \! \to \! {\rm SM}  \right) }
= \!\frac{c_{\! _R}^2   \, v_H^2}{4 s_{\! _L}^2 \, m_{t'}^2} \! \left( \! y_o^2 - 2y_oy_\chi s_{\! _L} \frac{m_{t'}}{m_t} \cos(\beta_o \! - \!\beta_\chi) + y_\chi^2 s_{\! _L}^2  \frac{m_{t'}^2}{m_t^2} \right) 	\!\! \left( 1 -  \frac{M_{a}^2}{m_{t'}^2}    \right)^{\! \! 2} \! .
\label{eq:Rt}
\ee
Depending on the values of the $s_{\! _L} m_t / m_{t'}$,  $y_o$ and $y_\chi$ parameters,  
the $R_t$ ratio may have any positive values from $R_t \gg 1$ ({\it i.e.}, $t'$ decays most of the time to $t  a_t $, or to $t  \varphi_t $ if kinematically allowed) to $R_t \ll 1$ ({\it i.e.}, $t'$ decays predominantly via standard modes). For example,  
\be
R_t  \approx   \frac{y_o^2}{4} \left(s_{\! _L}  \frac{m_{t'}}{m_t} \right)^{\! -2}  \! \left( 1 + s_{\! _L}^2 \, \frac{m_{t'}^2 }{m_t^2}  \right)^{\! -1} \; \; {\rm for} \;\;\; 
 s_{\! _L}  \frac{m_{t'}}{m_t}  \ll \frac{y_o}{y_\chi}  ~~,
 \label{eq:Rt-special}
\ee
which implies $R_t \gg 1$ for $s_{\! _L} m_{t'}/m_t \ll y_o$, and $R_t \ll 1$ for $s_{\! _L} \gtrsim y_o$. Recall that $y_o$ and $y_\chi$ are Yukawa couplings introduced in  (\ref{eq:Yukawas}), and may have any positive values below some unitarity bound of order one.
Some typical values for the parameters, $y_o \approx 1$, $m_{t'} \approx 1.5$ TeV,  
give $R_t \approx 3.4$ for $s_{\! _L} = 0.03$, 
and $R_t \approx 0.19$ for $s_{\! _L} = 0.1$;  note that the $t^\prime \! \to t  a_t$ mode dominates ($R_t > 1$) for $s_{\! _L} < 0.052$. In Sections~\ref{sec:eight} and \ref{sec:multi-lepton} we will focus on the parameter space in which the sum of branching fractions for the decays $t' \to t a_t$ and $t' \to t \varphi_t$ is close to 1 while the standard $t'$ decay modes are negligible. A parameter choice corresponding to this scenario and the associated branching fractions are shown in Table~\ref{tab:branchingfractions}.

Furthermore, a range of  values of the parameters gives $R_t$ of order one. For example, $y_o \approx 0.5$, $y_\chi \lesssim 0.2$, and $s_{\! _L}  m_t' / m_t$ increasing from  0.2 to 0.4 imply that  $R_t$ decreases from 1.5 to 0.34, with little sensitivity to the $\beta_o \! - \!\beta_\chi$ phase.  Consequently, the $t^\prime$ branching fractions  into $t a_t $ and into SM modes  are of the same order of magnitude. This motivates searches for mixed $t'$ decay topologies (referred to as hybrid signals in Section~\ref{sec:other}), in which $t^\prime \bar t^\prime $ production is followed by a $t^\prime \to t h^0 / t Z / b W$ decay and a $\bar t^\prime \to \bar t a_t $ decay, or the charge-conjugated process. Table~\ref{tab:branchingfractions} shows the branching fractions for a parameter point where the fraction of such hybrid signals among $t'\bar{t}'$ production events is close to maximal (see Section~\ref{sec:other} for a more detailed discussion).

\begin{table}[t!]
	\begin{center}
		\begin{tabular}{c|cccc}
			\toprule
			Parameters & $\mathcal{B}(t' \to t a_t)$ & $\mathcal{B}(t' \to t h^0)$ & $\mathcal{B}(t' \to t Z)$ & $\mathcal{B}(t' \to b W)$  \\\hline
			$y_o=2$, $s_L=0.03$  & 0.951 & 0.013 & 0.011 & 0.024 \\
			$y_o=1$, $s_L=0.05$ & 0.749 & 0.067 & 0.060 & 0.124 \\
			$y_o=0.7$, $s_L=0.05$  & 0.500 & 0.134 & 0.118 & 0.248  \\
			$y_o=0.7$, $s_L=0.1$  & 0.224 & 0.206 & 0.183 & 0.386  \\
			\bottomrule
		\end{tabular}
	\end{center}
	\caption{Branching fractions of $t'$ for a few parameter choices, based on the leading-order widths \eqref{eq:tprime_to_tat_width} and \eqref{eq:standard}. Here we set $y_\chi=y_o$, fixed the masses at $m_{t'}=1.5$~TeV, $M_a=400$~GeV, $M_\varphi>m_{t'}$, and set the complex phases to $\beta_o=0$, $\beta_\chi=\pi$. \label{tab:branchingfractions}}
\end{table}

Let us consider briefly the case where the mass terms (\ref{eq:mass-terms}) are not present, {\it i.e}, $m_{\chi\chi} = m_{\chi 3} = 0$. In that case, the $t'$ quark obtains its mass almost entirely 
from the VEV of $\phi$, and $a_t$ can be identified as the pseudo-Nambu-Goldstone boson associated with a global $U(1)$ symmetry that is explicitly broken by some suppressed terms in the $\phi$ potential. Then, as proven in \cite{Bernreuther:2023uxh} for a similar Lagrangian, 
\be
m_{t^\prime} \approx v_\phi \sqrt{y_{\chi}^2 + y_o^2} ~~,
\ee
and $R_t$ is determined by the ratio of the $H$ and $\phi$ VEVs:
\begin{equation}
	R_t \approx \frac{v_H^2}{4v_\phi^2} \approx  7.6 \times 10^{-3} \, \left(y_\chi^2 + y_o^2\right) \, \left(\frac{1~\mathrm{TeV}}{m_{t^\prime}}\right)^2 ~~,
\end{equation}
where we kept only the leading order in $s_{\! _L}^2$ and  $m_t^2/m_{t^\prime}^2$.
The unitarity requirement that $y_\chi,\, y_o \lesssim 2$ thus leads to an upper bound on $R_t$ in the Nambu-Goldstone limit, which imposes $R_t \lesssim 0.061$ for $m_{t^\prime} = 1$~TeV, and even more stringent bounds for larger $m_{t^\prime} $. Thus, the discovery mode for the $t'$ quark in the Nambu-Goldstone limit is given by the standard decays of pair-produced vectorlike quarks.

The main decays of the pseudoscalar $a_t$ proceed through the diagonal couplings to $t\bar t$ and $t' \bar t'$ given in  (\ref{eq:scalar_t_coupling}) and  (\ref{eq:scalar_tprime_coupling}). 
For the mass ordering relevant to the LHC phenomenology analyzed in Section~\ref{sec:8top}, $2 m_t < M_a < m_{t^\prime} - m_t$, the only tree-level decay mode of the pseudoscalar is $a_t \to t \bar{t}$ with partial width
\begin{equation}
	\Gamma \! \left( a_t \to t \bar{t} \right)  =  \frac{3 \, s_{\! _L}^2 c_{\! _R}^2 }{16 \pi }
	\left| {\cal Y}   \right|^2    M_a \left( 1 - 4 \frac{m_t^2}{M_a^2} \right)^{3/2}   ~~.
\label{eq:att}	
\end{equation}

In addition, $a_t$ can undergo loop-induced decays into gluons, as long as $M_a$ lies sufficiently far above $\Lambda_\mathrm{QCD}$ that gluons are the appropriate degrees of freedom. The decay can proceed through a top quark loop due to the coupling in (\ref{eq:scalar_t_coupling}), or through a $t^\prime$ loop due to the coupling in (\ref{eq:scalar_tprime_coupling}), and has a partial width  
\begin{equation}  \hspace*{-0.3cm}
	\Gamma \! \left( a_t \to g g \right) = 
	\frac{ \alpha_s^2  \, c_{\! _R}^2  M_a^3 }{ 64\pi^3 \, m_{t^\prime}^2} \,   \left[ 
	\left( {\rm Re} \, {\cal Y}^{\,\prime}  \right)^2 \!
	+ \frac{4}{9} \left( {\rm Im} \, {\cal Y}^{\,\prime}   \right)^2   + 
s_{\! _L}^2  \, \frac{m_{t'}^2}{m_t^2}   \left( \left( {\rm Re} \,  {\cal Y} \right)^2 \!
	+ \frac{4}{9} \left( {\rm Im} \,  {\cal Y} \right)^2  \right)   \right]   ,
\end{equation}
up to corrections of order $M_a^2/(2m_{t^\prime})^2$. The strong coupling constant $\alpha_s$ is evaluated at the $M_a$ scale, and it varies from  $\alpha_s \approx 0.098$ for $M_a = 0.4$ TeV, to $\alpha_s \approx 0.089$ for $M_a = 1$ TeV. 
For $M_a > 2 m_t$, the branching fraction for $a_t \to g g $ is typically below a few percent, because the competing mode $a_t \to t \bar{t}$ is not loop suppressed. However, 
(\ref{eq:att}) shows that $\Gamma \left( a_t \to t \bar{t} \right)$ is suppressed by $s_{\! _L}^2$, so for sufficiently small $s_{\! _L} \ll 1$ it may be possible that $a_t \to g g$ is the dominant decay mode.
Note that the effective Yukawa couplings ${\cal Y}$ and ${\cal Y}^{\,\prime}$, defined in  (\ref{eq:effYuk}) and  (\ref{eq:effYukprime}),
do not generically vanish in the $s_{\! _L} \to 0$ limit. 

To estimate the values of  $s_{\! _L} \ll  m_t/m_{t'} $ where the $a_t \to g g$ branching fraction becomes comparable to the $a_t \to t \bar{t}$ one, let us set for simplicity $y_o \approx y_\chi$, so that the ratio of widths becomes
\be 
	R_g \equiv \frac{\Gamma(a_t \to gg)}{\Gamma(a_t \to t \bar{t} \, )} \approx  
	\frac{\alpha_s^2 \, M_a^2 }{6 \pi^2 \, m_{t^\prime}^2  \,s_{\! _L}^2 } 
\left( 1 - 4 \frac{m_t^2}{M_a^2} \right)^{\!-3/2}   ~~.
\label{eq:Bratio}
\ee 
Taking for illustration some typical mass ranges, $M_a \in [0.4,1]$ TeV, $m_{t^\prime} \in [1.5,2]$ TeV,  we find $R_g <1$ when $s_{\! _L}$ is larger than certain values within the range  $(5.1-9.3)\times 10^{-3}$, while $s_{\! _L} > 0.02$ ensures $R_g < 0.2$, corresponding to a branching fraction ${\cal B}(a_t \to t \bar{t} \, ) > 83\%$.

The scalar $\varphi_t$, like $a_t$, can decay at tree level into $t\bar{t}$, with a partial width suppressed by $s_{\! _L}^2$, or at one loop into $gg$. 
However, if $M_\varphi > 2 M_a$, there exists an additional 2-body decay mode $\varphi_t \to a_t a_t$, which occurs due to interactions in the $\phi$ potential and is neither mixing- nor loop-suppressed. Note that even a stronger mass hierarchy, $M_a \ll M_\varphi$, would be  well-motivated since $a_t$ is a pseudo-Nambu-Goldstone boson, while $\varphi_t$ is not.
Hence, $\varphi_t \to a_t a_t$ is expected to be the dominant decay mode of the scalar. 

Further suppressed decay modes of $a_t$ and $\varphi_t$ include $\gamma\gamma$, $Z\gamma$, $ZZ$, and $W^+W^-$,  
which occur at one loop through a sum of triangle diagrams that involve the $t$, $t'$ and $b$ quarks. The $t' \to t a_t \to t \gamma\gamma$ mode 
is shown in \cite{Wang:2020ips} to be promising due to small backgrounds. There are also 3-body decays at tree level, such as  $\varphi_t \to \bar t \,  t^{\prime *} \to \bar t t a_t $, but which  are unlikely to provide discovery modes.

Additional channels are open in the presence of a Higgs portal interaction, $\phi^\dagger  \phi H^\dagger H$; here we will assume that its coupling is small enough to be inconsequential. Single production of $a_t$ (or $\varphi_t$) at the LHC through gluon fusion is suppressed either by $s_{\! _L}^2$ in the case of the top loop [because of the coupling (\ref{eq:scalar_t_coupling})], or by the small ratio $(M_a/m_{t'})^2$ in the case of the $t'$ loop.
Thus, single $a_t$ production, followed by $a_t \to t\bar t$, cannot be probed in Run 2 or Run 3 of the LHC  (for a study of sensitivity below the  $t\bar t$ threshold, see \cite{Blasi:2023hvb}).

Production of $a_t$ in association with a $t\bar t$ pair is also suppressed by $s_{\! _L}^2$. As a result the process 
\be
pp \to  a_t \, t \,  \bar t    \to   ( t \bar t )  \,  t \, \bar t    
 \label{eq:tta}
\ee
has a leading-order cross section (computed with \textsc{MadGraph} \cite{Alwall:2014hca}) of only $0.20$ fb at $\sqrt{s} = 13$ TeV, and $0.23$ fb at $\sqrt{s} = 13.6$ TeV, for $M_a = 400$ GeV and a nearly maximal coupling in (\ref{eq:scalar_t_coupling}): 
$s_{\! _L} c_{\! _R} {\cal Y} \approx 8.6 \times 10^{-2}$, which is obtained for  $s_L = 0.1$, $y_o = 1$, $y_\chi \ll 1$, $m_{t'} = 1$ TeV.
For comparison, the current uncertainty in the measurement of the 4-top process is 4 fb \cite{CMS:2023ftu,ATLAS:2023ajo}, so the process (\ref{eq:tta})
cannot set yet  useful constraints for $a_t$ with mass above the $t \bar t$ threshold. Nevertheless, searches for the process (\ref{eq:tta}) at the High-Luminosity LHC runs may take advantage of the fact that two of the four tops form a resonance of mass $M_a$.


\section{LHC signals of the $t'$ quark}  
\label{sec:8top}
\setcounter{equation}{0}

At the LHC, pair production of $t'$ quarks has a sizable and model-independent cross section, as it proceeds through the QCD coupling of gluons to quarks, and thus depends only on the $t'$ mass, $m_{t^\prime}$.  Single $t'$ production is highly suppressed, either by $s_{\! _L}^2$ when it is produced via its couplings to SM bosons [{\it e.g}, see (\ref{eq:W-offdiagonal})], or by phase space when it is produced  in association with the new spin-0 particles through the coupling (\ref{eq:scalar_t_tprime_coupling}).

\begin{figure}[b!]
	\begin{center}
		\includegraphics[width=0.73\textwidth]{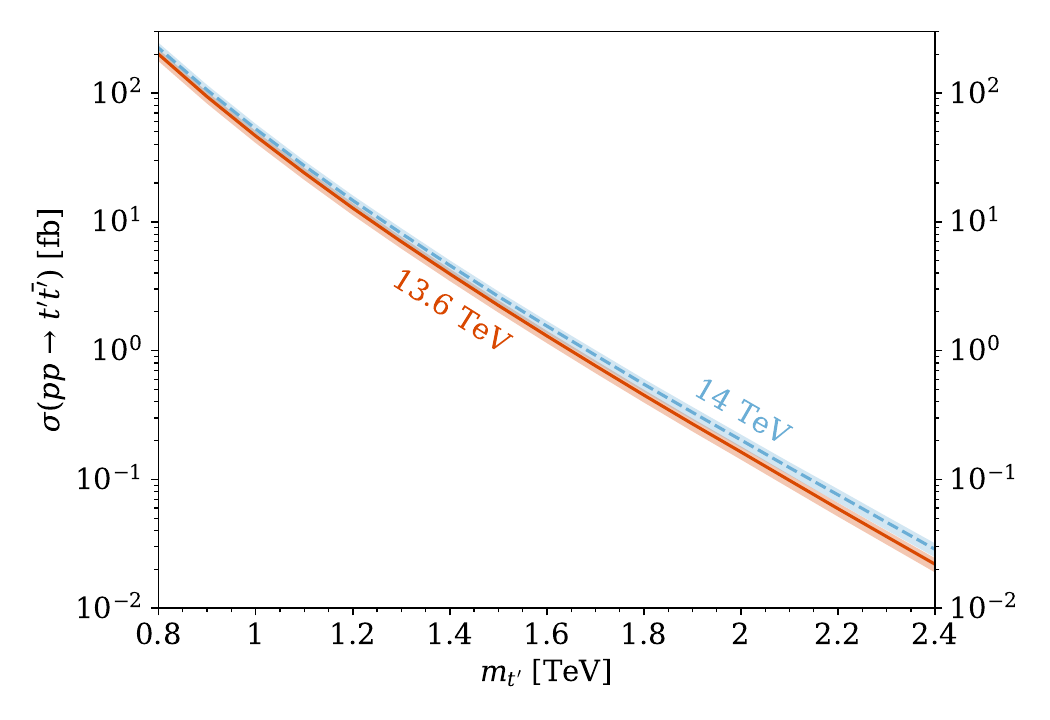}
\vspace*{-0.3cm}		
		\caption{\label{fig:nlo_xsec} 
		Cross section for $t' \bar{t}'$ production in proton-proton collisions, computed at NLO with \textsc{MadGraph}~\cite{Alwall:2014hca}  at  a center-of-mass energy of  
		$13.6$~TeV (solid red line) or $14$~TeV (dashed blue line). The scale uncertainty is indicated by envelopes around each line (see main text for details).}
	\end{center}
\end{figure}

We calculate the total cross section for the process $p p \to t' \bar{t}'$ at next-to-leading order (NLO) with \textsc{MadGraph}~\cite{Alwall:2014hca}
(version  \textsc{MadGraph5\_aMC@NLO}~3.4.2) using the \textsc{NNPDF23\_nlo\_as\_0118\_qed} PDF set~\cite{Ball:2013hta}. The renormalization and factorization scales are set to the sum of the transverse masses of the $t'$ quarks and initial-state partons divided by two. The resulting NLO cross sections at $\sqrt{s}=13.6$~TeV and $\sqrt{s}=14$~TeV are shown in Figure~\ref{fig:nlo_xsec} as a function of $m_{t'}$. Envelopes around the lines in Figure~\ref{fig:nlo_xsec} indicate the scale uncertainty obtained by varying the renormalization and factorization scale within a factor of two in either direction. A comparison of the NLO cross sections to the leading-order ones can be found in \cite{Fuks:2016ftf}.

In the current Run 3, with $\sqrt{s}=13.6$~TeV, the production cross section is already large enough so that more than 30 events with  $t' \bar{t}'$ can be produced in 300 fb$^{-1}$ of data if  
 $m_{t'} = 2$ TeV. The number of events grows to about 400  at the High-Luminosity LHC,  with $\sqrt{s}=14$~TeV and 3000  fb$^{-1}$ of data. Thus, the LHC experiments will potentially  be sensitive to various decay modes of the vectorlike quark.

\subsection{Signals with 6, 8, or 10  top quarks}
\label{sec:eight}

The different decay modes of the three new particles introduced in Section~\ref{sec:model} (the $t^\prime$ quark, the $a_t$ pseudoscalar, and the $\varphi_t$ scalar) lead to several possible signatures of $t^\prime \bar t^\prime  $ production at the LHC.
The total branching fraction of the $t'$ quark into particles beyond the SM, 
\be
\mathcal{B}_{t^\prime \to a/\varphi} \equiv \mathcal{B}(t^\prime \to t a_t)+\mathcal{B}(t^\prime \to t \varphi_t) ~~,
\ee 
can vary widely, as follows from (\ref{eq:Rt}) and (\ref{eq:Rt-special}). Hence, it is useful to treat the branching fraction $\mathcal{B}_{t^\prime \to a/\varphi}$ as a phenomenological parameter and organize the signatures according to its value.
For $\mathcal{B}_{t^\prime \to a/\varphi} \ll 1$, the standard decay channels of $t^\prime$ dominate, so each $t^\prime$ produced at the LHC decays to either $t h$, $t Z$ or $b W$. The ensuing signals have been searched for by the ATLAS \cite{ATLAS:2024gyc} and CMS \cite{CMS:2022fck} collaborations, which have excluded $m_{t'} \lesssim 1.4$~TeV in this standard scenario.

When $\mathcal{B}_{t^\prime \to a/\varphi}$ is larger than about 0.7,  the main signals of $t^\prime \bar t^\prime$ production arise from $t^\prime$ decaying to $t a_t$ or $t \varphi_t$,  and the corresponding $\bar t^\prime$ decays. In the following we will focus mainly on the part of parameter space where the decay $a_t \to t\bar{t}$ dominates over $a_t \to gg$, which is the case for sufficiently large mixing angle,  $s_L \gtrsim 10^{-2}$, as follows from \eqref{eq:Bratio}. The phenomenology associated with $a_t \to gg$ is discussed, for example, in \cite{Cacciapaglia:2019zmj} (see also \cite{Dobrescu:2016pda} for other $t\bar t + 4j$ signals of vectorlike quarks). 
As discussed in Section~\ref{sec:mass},  the dominant decay mode of the scalar is likely $\varphi_t \to a_t a_t$. With the subsequent decay $a_t \to t\bar{t}$, each $\varphi_t$ thus produces two $t\bar t$ pairs.

\begin{figure}[t]
	\begin{center}
		\includegraphics[width=0.45\textwidth]{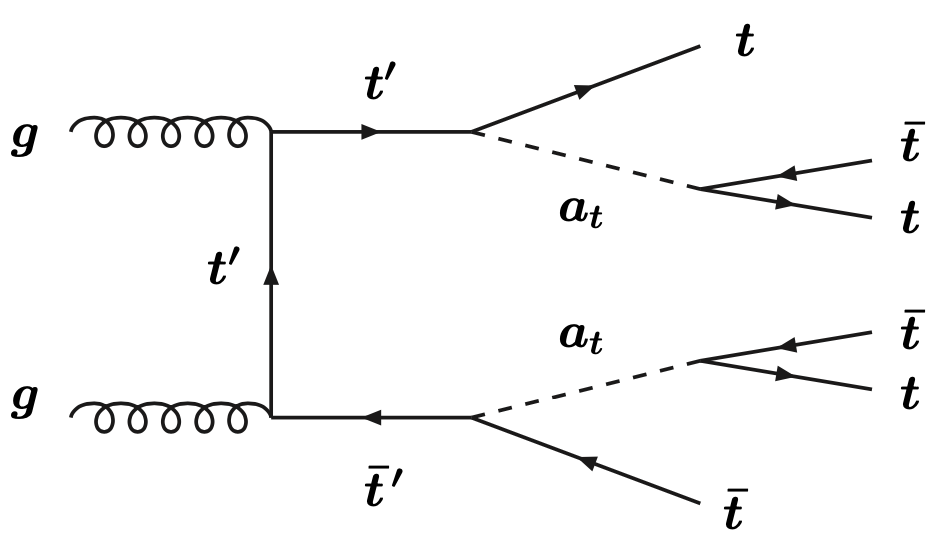}
		\hspace{8mm}
		\includegraphics[width=0.45\textwidth]{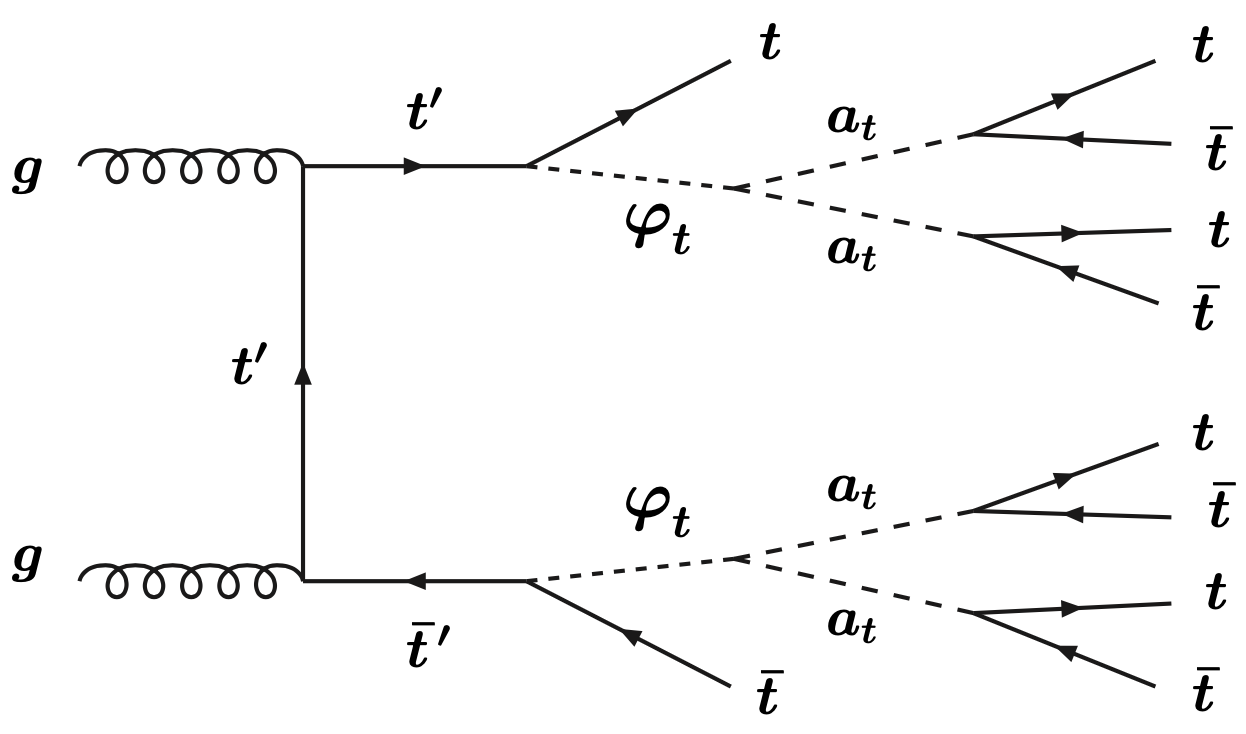} 
		\caption{\label{fig:diagram6t_10t}$t'\bar{t}'$ production at the LHC leading to a 6-top (left) or 10-top (right) final state. 
		}
	\end{center}
\end{figure}

\begin{figure}[t]
	\begin{center}
		\hspace{-8mm} \includegraphics[width=0.6\textwidth]{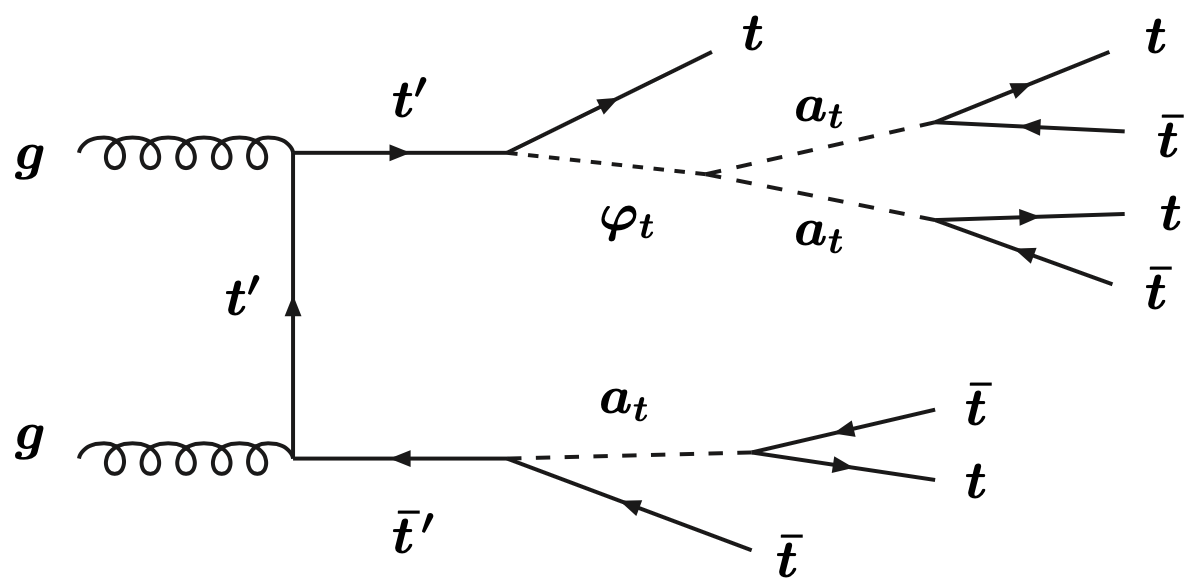} 
		\caption{\label{fig:diagram8t}$t'\bar{t}'$ production at the LHC leading to an 8-top final state. 
		}
	\end{center}
\end{figure}

An event with a $t'\bar{t'}$ pair followed by $t' \to ta_t$ as well as $\bar t' \to \bar{t}a_t$ gives rise to a ``6-top'' signal (three $t\bar{t}$ pairs) as depicted in the left panel of Figure~\ref{fig:diagram6t_10t}. If $t'$  and $\bar{t}'$ instead decay to $t \varphi_t$ and $\bar{t}\varphi_t$, respectively, the result is a spectacular ``10-top'' signal (five $t\bar{t}$ pairs) as shown in the right panel of Figure~\ref{fig:diagram6t_10t}. Furthermore, $t^\prime \bar t^\prime$  events can be of the mixed-decay type,
\begin{equation}
	pp \to t^\prime \bar t^\prime \to  (a_t\varphi_t) t \bar t \to (3 a_t) t\bar t \to 4 (t \bar t) ~,
\end{equation}
which we refer to as ``8-top" events. This type of event is displayed in the diagram of Figure~\ref{fig:diagram8t}.

The relative fractions of 6-top, 8-top and 10-top events are proportional to the combined branching fractions  $\mathcal{B}(t^\prime \to t a_t)^2$, $2\,\mathcal{B}(t^\prime \to t a_t)\,\mathcal{B}(t^\prime \to t \varphi_t)$ and $\mathcal{B}(t^\prime \to t \varphi_t)^2$, respectively, assuming that $\mathcal{B}(\varphi_t \to a_t a_t)$ is close to 100\%. Since the couplings of $a_t$ and $\varphi_t$ to $\bar{t} t'$ (and $\bar{t}' t$) given in \eqref{eq:scalar_t_tprime_coupling} differ merely by a complex phase, the ratio of branching fractions for the decays $t' \to t \varphi_t$ and $t' \to t a_t$ depends only on the mass ratios, 
as dictated by the matrix elements and the available phase space:
\be
\frac{\mathcal{B}(t^\prime \to t \varphi_t)}{\mathcal{B}(t^\prime \to t a_t)} 
= \frac{f(M_\varphi/m_{t^\prime}, \, m_t/m_{t^\prime})}{f(M_a/m_{t^\prime}, \, m_t/m_{t^\prime})}
\approx \left(\frac{m_{t^\prime}^2-M_\varphi^2}{m_{t^\prime}^2-M_a^2}\right)^{\! 2} \; ,
\label{eq:phia}
\ee
where the function $f$ is defined as
\be
f(x,y) = \left( 1 - x^2 - y^2 \right) \left( (1 - x^2)^2 - y^2 (2 + 2 x^2 - y^2) \rule{0mm}{3.9mm} \right)^{\! 1/2}  ~~.
\ee
Note that $f(x, 1- x) = 0$ as the 2-body phase space vanishes for  $M_\varphi = m_{t^\prime} - m_t$.
In the last step of (\ref{eq:phia}) we neglected all terms dependent on $m_t^2$, which is a good approximation only if $M_\varphi$ is not very close to the 
$t^\prime \to t \varphi_t$ threshold.
Since $\mathcal{B}(t^\prime \to t \varphi_t) < \mathcal{B}(t^\prime \to t a_t)$ for any $M_\varphi > M_a$, the 10-top signal always has the smallest rate. However, whether the 8-top signal has a larger rate than the 6-top signal depends on the particle masses, because of the extra factor of 2 in the combined branching fraction for the 8-top signal, which compensates for the smaller $\mathcal{B}(t^\prime \to t \varphi_t)$.

\begin{figure}[t!]
	\begin{center}
		\includegraphics[width=0.72\textwidth]{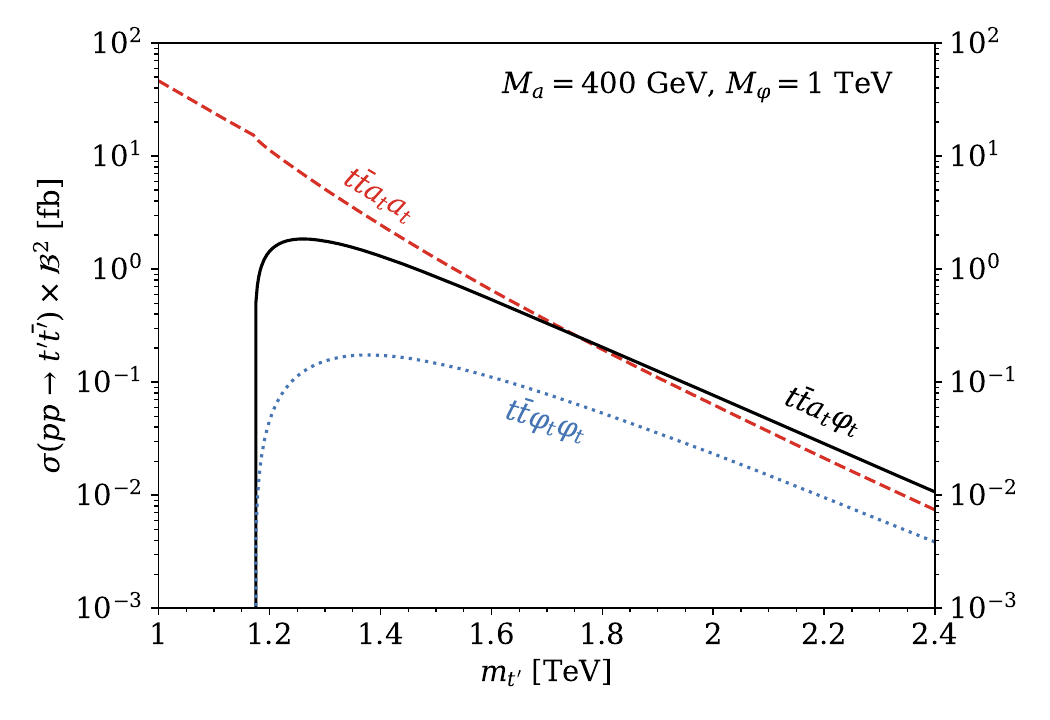}
		\caption{\label{fig:xsecxbr}Cross section for $t' \bar t'$ production at the 13.6 TeV LHC multiplied by the branching fractions  for both $t'$ and $\bar{t}'$ to decay into a top quark and a new spin-0 particle, which can be either the pseudoscalar $a_t$ or the scalar $\varphi_t$. The possible final states are $t\bar t a_t  a_t \to 3(t\bar t)$ (dashed red line), $t\bar t \, a_t  \varphi_t \to t \bar t \, a_t  a_t  a_t  \to 4(t\bar t)$ (solid  black line), or $t\bar t \, \varphi_t  \varphi_t \to t \bar t \, a_t  a_t  a_t  a_t \to 5(t\bar t)$  (dotted blue line). The $a_t$ and  $\varphi_t$ masses are fixed at 400~GeV and 1~TeV, respectively, and $\mathcal{B}(\varphi_t \to a_t a_t)$ is taken to be 1.}
	\end{center}
\end{figure}

The production cross sections for $t \bar{t} a_t a_t$ (6-top), $t \bar{t} a_t \varphi_t$ (8-top),  and $t \bar{t} \varphi_t \varphi_t$ (10-top) events are shown in Figure~\ref{fig:xsecxbr} for $M_a=400$~GeV and $M_\varphi=1$~TeV. The mixing-induced decays of $t'$ directly into SM particles are assumed here to have negligible widths. For $m_{t'} < M_\varphi + m_t \approx 1.2$~TeV, the only possible 2-body decay mode of $t'$ is into $t a_t$, so that $t \bar{t} a_t a_t$ is the only non-negligible signal of $t'\bar{t}'$ production. The corresponding 6-top signal leads to a signature of up to three same-sign leptons and up to six $b$-tagged jets. Existing constraints on this signal arise from LHC searches for final states with many tops or for jets in combination with (same-sign) leptons. In Ref.~\cite{Han:2018hcu}, the ATLAS searches of Refs.~\cite{ATLAS:2016dlg, TheATLAScollaboration:2016gxs} were recast for a simplified $t'$ model and shown to exclude $m_{t'} \lesssim 1$~TeV in a scenario where $t'$ pair production results exclusively in 6-top events.

For $m_{t'} > M_\varphi + m_t$ the decay channel $t' \to t \varphi_t$ opens up, but $t \bar{t} a_t a_t$ remains the dominant signal process for larger $m_{t'}$ (up to $m_{t'} \approx 1.8$~TeV for the parameters used in Figure~\ref{fig:xsecxbr}) due to the smaller branching fraction for  $t' \to t \varphi_t$. For $m_{t'} \gtrsim 1.8$~TeV, $t \bar{t} a_t \varphi_t$ becomes the dominant signal process, resulting in 8-top events. The corresponding cross section is about 0.3~fb at $m_{t'} = 1.8$~TeV, so that the number of 8-top events that can be produced at that mass in Run 3 of the LHC is $\sim \!10^2$.  The 10-top signal is subdominant, as expected, but can still reach a cross section of up to 0.2~fb for $M_\varphi=1$~TeV and $M_a=400$~GeV.

\begin{table}[b!]
	\begin{center}
		\begin{tabular}{c|ccccccc}
			\toprule
		    \backslashbox{Process}{Leptons} & 2 SS & 2+1 & 3 SS & 2+2 & 3+1 & 4 SS & $\geq5$ \\\hline
			6 top  & 0.12 & 0.13 & 0.015 & 0.023 & 0.016 &  0 & 0.0059 \\
			8 top  & 0.13 & 0.19 & 0.032 & 0.051 & 0.045 & 0.0028 & 0.034 \\
			10 top  & 0.12 & 0.21 & 0.043 & 0.076 & 0.076 & 0.0076 & 0.094 \\
			\bottomrule
	\end{tabular}
	\end{center}
	\caption{\label{tab:lepton_signatures}Fraction of signal events (before event selection)  that have two leptons of same-sign charge (``2SS"), three leptons with mixed charges (``2+1"), three same-sign leptons (``3SS"), two pairs of opposite-sign leptons (``2+2"), three same-sign  leptons plus a lepton of opposite charge (``3+1"), four same-sign  leptons (``4SS"), and five or more leptons (``$\geq 5$"). Each category is shown separately for $3 (t \bar t)$ (``6-top"),  $4 (t \bar t)$ (``8-top"), and $5 (t \bar t)$ (``10-top") processes.}
\end{table}

\subsection{Multi-lepton signatures}
\label{sec:multi-lepton}

While reconstructing such a large number of top quarks is challenging, other experimental handles on these signals exist due to the potentially large number of (same-sign) leptons in the final state from top decays. Table~\ref{tab:lepton_signatures} lists the fractions of 6-top, 8-top, and 10-top events that result in various interesting lepton signatures, which may serve to distinguish signal from background. These include two same-sign leptons, three leptons with mixed charges, three same-sign leptons, two pairs of opposite-sign leptons, three same-sign leptons combined with one lepton of opposite charge, four same-sign leptons, and five or more leptons. As expected, 8-top and 10-top events typically have higher lepton multiplicity than 6-top events.

\begin{figure}[t!]
	\begin{center}
		\hspace*{-0.25cm}\includegraphics[width=0.533\textwidth]{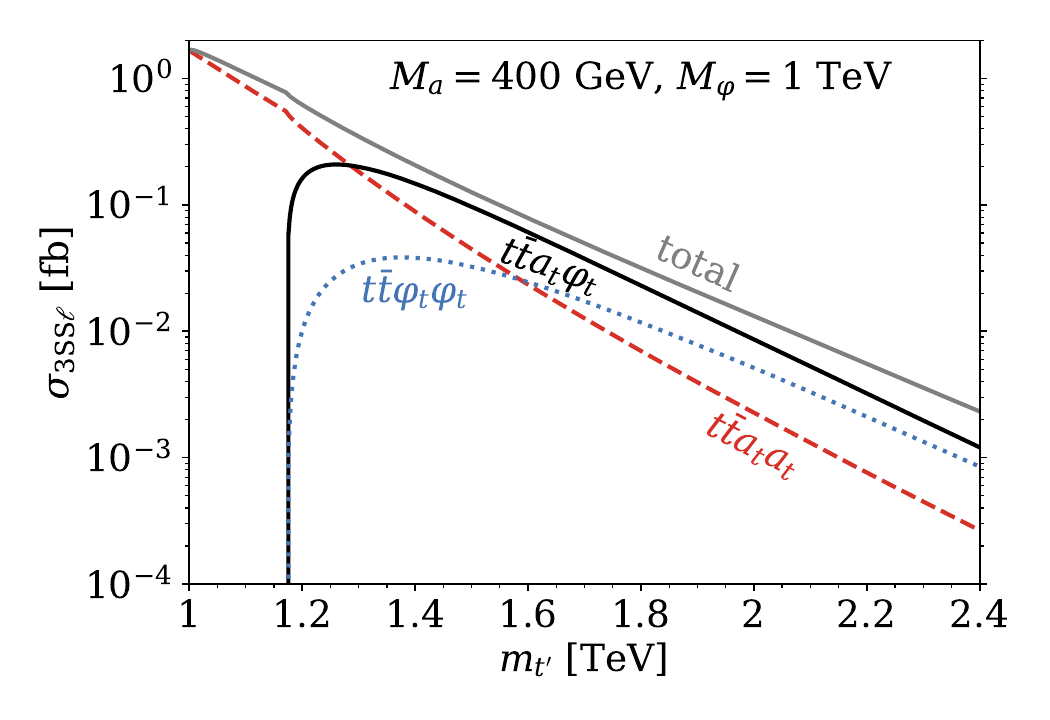}\hspace*{-0.55cm}
		\includegraphics[width=0.533\textwidth]{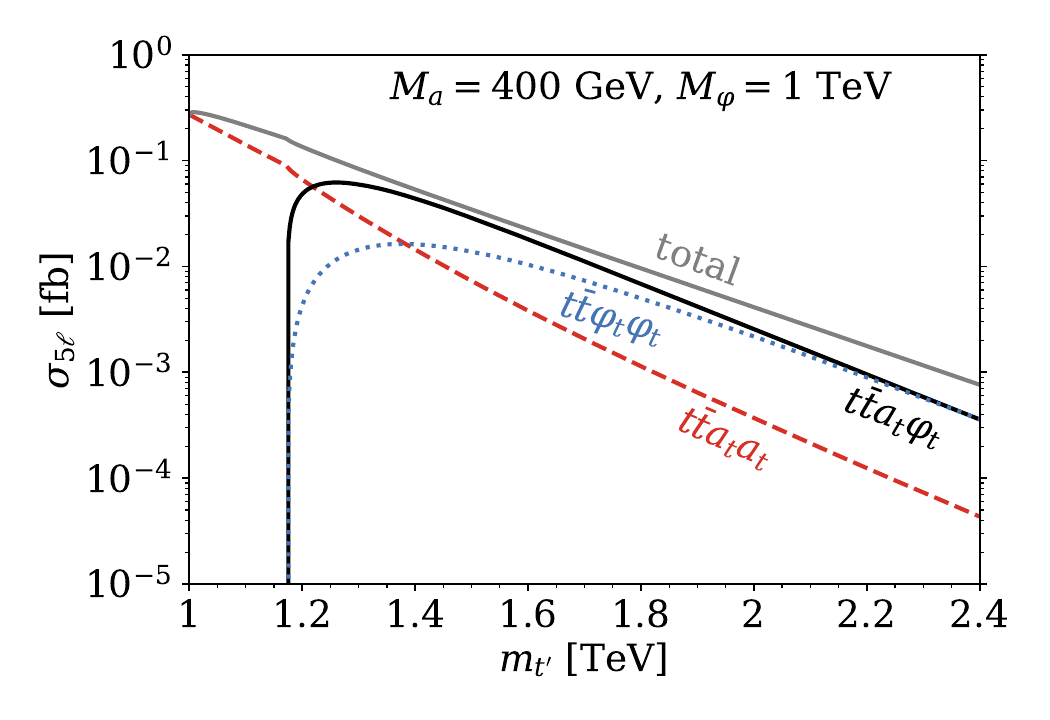}
		\caption{\label{fig:xsec_lepton_signatures}Branching fractions times the $pp \to t' \bar t'$ NLO cross section at $\sqrt{s}=13.6$~TeV, as a function of the vectorlike quark mass,  for processes that produce 3 same-sign leptons (left panel) or 5 or more leptons (right panel). Solid gray lines represent the sum of the processes with $ t' \bar t' \to 3 (t \bar t)$ (red dashed lines), $4 (t \bar t)$ (solid black lines), and $5 (t \bar t)$ (dotted blue lines).
			The $a_t$ and  $\varphi_t$ masses are fixed at 400 GeV and 1 TeV, respectively.}
	\end{center}
\end{figure}

While Table~\ref{tab:lepton_signatures} shows the fraction of 6-top, 8-top, and 10-top events resulting in a particular signature, it does not take the differences in the production cross sections for each type of signal event into account. The left panel of Figure~\ref{fig:xsec_lepton_signatures} shows the total cross section for signal events with three or more same-sign leptons (3SS$\ell$) for $M_a=400$~GeV and $M_\varphi=1$~TeV, as well as a breakdown by signal process. The 8-top process from $t \bar{t} a_t \varphi_t$ production becomes the largest contribution to 3SS$\ell$ events for $m_{t'} \gtrsim 1.3$~TeV, only slightly above the kinematic threshold for $t' \to \varphi_t t$. With the combined contributions from all signal processes, the model is expected to yield some number of 3SS$\ell$ events in 300~fb$^{-1}$ of data for $t'$ masses of up to 2.4~TeV. The right panel of Figure~\ref{fig:xsec_lepton_signatures} shows the analogous cross sections for events containing five or more leptons. Again, 8-top events are the dominant contribution to this signature for almost any mass above the kinematic threshold for $t' \to \varphi_t t$.

\begin{figure}[t!]
	\begin{center}
		\includegraphics[width=0.515\textwidth]{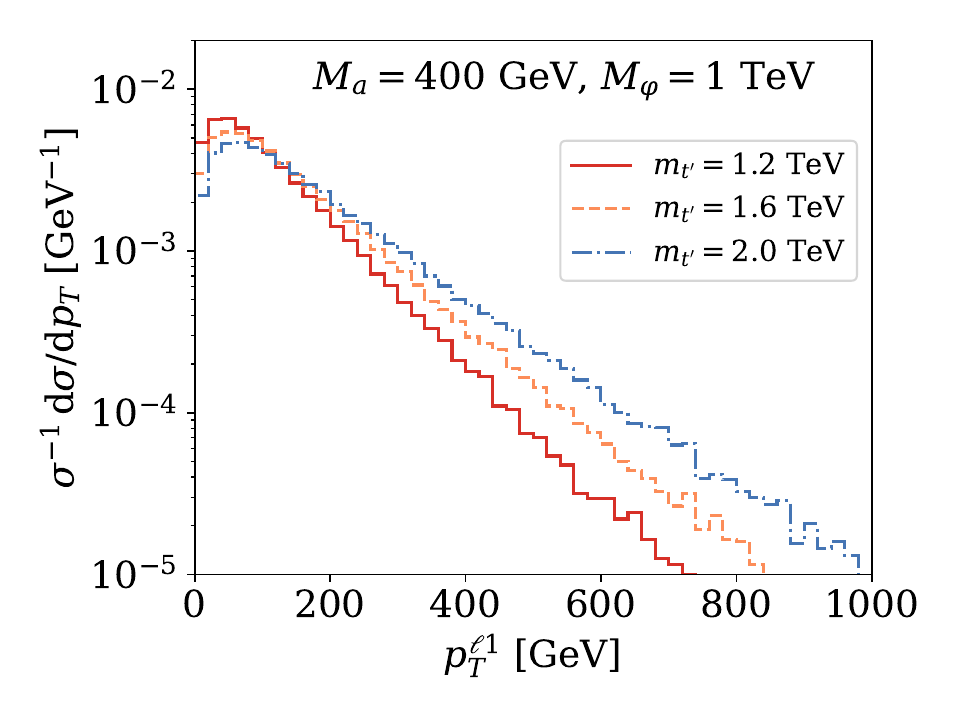}\hspace{-0.7cm}
		\includegraphics[width=0.515\textwidth]{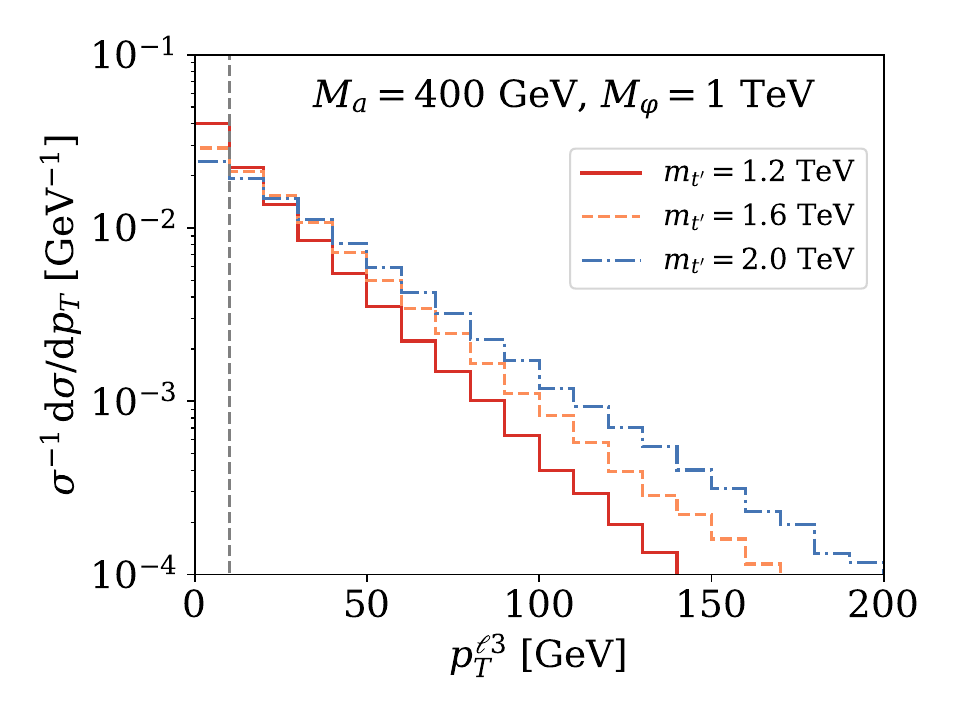}
		\caption{Distributions of the $p_T$ of the highest-$p_T$ lepton (left) and third-highest-$p_T$ lepton (right) in signal events at $\sqrt{s}=13.6$~TeV for different $t'$ masses. The signal consists of 6-top, 8-top, and 10-top events from $t\bar{t}a_ta_t$, $t\bar{t}a_t\varphi_t$, and $t\bar{t}\varphi_t\varphi_t$ production, respectively, with corresponding cross sections as shown in Figure~\ref{fig:xsecxbr}. \label{fig:lepton_pt}}
	\end{center}
\end{figure}

While the $t'$ quarks are heavy, the energy in 6-top, 8-top or 10-top events is distributed across a large number of final-state particles. Hence, it is a potential concern that leptons originating from these processes may be too soft to be detected.

To investigate the lepton kinematics, we generate signal events (consisting of the 6 top, 8 top, and 10 top contributions discussed in this Section) at $\sqrt{s}=13.6$~TeV with \textsc{MadGraph5\_aMC@NLO}~3.4.2~\cite{Alwall:2014hca} at LO using the \textsc{NNPDF23\_lo\_as\_0130\_qed} PDF set~\cite{Ball:2013hta} and a UFO model implemented in \textsc{FeynRules}~\cite{Alloul:2013bka}. We subsequently pass the events to \textsc{Pythia}~8.3~\cite{Sjostrand:2014zea} for parton showering and hadronization. The decay $a_t \to t \bar{t}$ is part of the \textsc{MadGraph} simulation, while the decay of the (anti-)top quarks via their leptonic and hadronic modes occurs in \textsc{Pythia}. We do not perform a detector simulation.

The resulting $p_T$ distributions of the highest-$p_T$ lepton and the third-highest-$p_T$ lepton in each event are shown in Figure~\ref{fig:lepton_pt} for $M_a=400$~GeV, $M_\varphi=1$~TeV, and $m_{t'}=1.2$~TeV, $1.6$~TeV, or $2$~TeV. As expected, based on the high multiplicity of the involved final states, the $p_T$ distribution of the leading (\emph{i.e.}\ highest-$p_T$) lepton peaks far below the $t'$ mass, between $p_T\approx 40$~GeV and $p_T\approx 60$~GeV for $m_{t'}$ between $1.2$~TeV and $2$~TeV. Still, the distribution remains sizable up to transverse momenta of several hundred GeV.

The analogous distributions for the third-highest-$p_T$ lepton are considerably softer. It is interesting to compare these to the minimum $p_T$ threshold implemented in searches for new physics in multi-lepton events. An example is the CMS analysis of Ref.~\cite{CMS:2022nty} which considers only leptons with $p_T>10$~GeV. Based on the distributions shown in the right panel of Figure~\ref{fig:lepton_pt}, we find that in 60\% of events with at least three leptons all three leading leptons have a $p_T>10$~GeV when $m_{t'}=1.2$~TeV. This fraction increases to 76\% for $m_{t'}=2$~TeV.

Furthermore, we find that the leptons in these events are mainly produced centrally, with more than 99\% of leading and more than 95\% of third-highest-$p_T$ leptons having pseudorapidity $|\eta|<2.4$ for any $m_{t'} > 1.2$~TeV.

Finally, let us comment briefly on possible backgrounds to the signals discussed here. $t\bar{t}t\bar{t}$ production, whose LHC cross section has recently been measured by ATLAS and CMS to be $22.5\substack{+6.6 \\ -5.5}$~fb and $17.7\substack{+4.4 \\ -4.0}$~fb, respectively, at $\sqrt{s}=13$~TeV~\cite{ATLAS:2023ajo, CMS:2023ftu}, can lead to up to four $b$-jets in association with up to four leptons. Moreover, a 3SS$\ell$ signature can arise from this process if the charge of one lepton is misidentified. Backgrounds that do not require charge misidentification to produce a 3SS$\ell$ signature, but which lead to fewer $b$-jets, are $t\bar{t}W^\pm W^\pm jj$ and $W^+W^-W^\pm W^\pm$. The production cross section for both of these processes is approximately $2\times 10^{-2}$~fb at $\sqrt{s}=13.6$~TeV. Other relevant backgrounds producing $b$-jets together with the various lepton signatures in Table~\ref{tab:lepton_signatures} include $t\bar{t}W^+W^-$, $t\bar{t}W^\pm$, and $t\bar{t}Z$, all of which can also have additional jets in the final state.

\bigskip

\subsection{Hybrid $t'$ signals}
\label{sec:other}

As discussed in Section~\ref{sec:mass}, there exists a sizable region of parameter space where the $t'$ branching fractions into $t a_t$ and 
$t \varphi_t$ are comparable to those into standard modes. This motivates the study of `hybrid' signals, where $t' \bar t'$ production is followed by 
one heavy quark decaying into a top quark plus one of the new spin-0 particles, and the other one decaying into $b W$, $t Z$ or $t h^0$ (or the charge-conjugate modes). The cascade decays
\be
pp \to t' \bar t'  \ba{c} \nearrow   \\ [-1mm] \searrow  \ea  \!\!\!
\ba{c}  (t a_t)(\bar b W^-)  \to  (t t \bar t )(\bar b W^-)  \\ [4mm]   (b W^+)(\bar t a_t)  \to  (b W^+)(\bar t t \bar t )   \ea  
 \ba{c} \searrow   \\ [0mm] \nearrow  \ea  \;  4(b W)    ~~
 \label{eq:Wa}
\ee
lead to final states with four $W$ bosons and four $b$-jets.  Thus, this type of hybrid signal (which is also
mentioned in \cite{Bhardwaj:2022nko}) is similar to the SM 4-top process, with the important difference that one of the $bW$ systems has an invariant mass (given by $m_{t'}$) that is about three times larger than the other three $b W$ systems.

In the optimal case for these final states, the $t a_t$ branching fraction satisfies
\be
\mathcal{B}(t^\prime \to t a_t) \approx  2 \mathcal{B}(t^\prime \to b W) \approx  4\mathcal{B}(t^\prime \to t Z) \approx 4\mathcal{B}(t^\prime \to t h^0) ~,
\ee
and $\mathcal{B}(t^\prime \to t \varphi_t) = 0$ (for $\varphi_t$ heavier than $t'$), 
so that the combined branching fraction of the cascade decays (\ref{eq:Wa}) reaches its maximum at $2(1/2)(1/4) = 25\%$. Thus, it is preferable for searches of hybrid signals to include cascade decays involving a $Z$ or Higgs boson:
\be
pp \to t' \bar t'  \ba{c} \nearrow   \\ [-1mm] \searrow  \ea  \!\!\!
\ba{c}  (t a_t)(\bar t + Z/h^0)  \to  (t t \bar t )(\bar t + Z/h^0)  \\ [4mm]   (t + Z/h^0)(\bar t a_t)  \to  (t + Z/h^0)(\bar t t \bar t )   \ea  
 \ba{c} \searrow   \\ [0mm] \nearrow  \ea  \;  4(bW) +  Z/h^0   ~~.
  \label{eq:Za}
\ee
The top quark and the $Z$ or Higgs boson arising from one top-prime decay have large boosts, given roughly  by $m_{t'}/(2m_t)$, which grows from 3 to 6 for $m_{t'}$ in the $1-2$ TeV range. Thus, dedicated taggers for boosted top and boosted $Z$ or $h^0$ can suppress the backgrounds for fully-hadronic  $t' \to t + Z/h^0$ final states, which have the largest branching fractions. The other top-prime cascade decays, $\bar t' \to \bar t t \bar t \to 3(bW)$ involves smaller boosts due to the higher multiplicity of the final state, so it may be better for first searches of this type to focus on leptonic decays of one or two $W$ bosons.

Depending on the mass ordering, there may also be hybrid signals arising from processes similar to those in (\ref{eq:Wa}) and (\ref{eq:Za}) with the $a_t$ pseudoscalar replaced by the $\varphi_t$ scalar. The main difference in that case is an additional $t\bar t$ pair in the final states, arising from $\varphi_t \to a_t a_t \to (t\bar t)(t\bar t)$. This leads to higher multiplicities, but also to lower-$p_T$'s of the $W$'s and $b$ jets not associated with standard $t'$ decays.

The measurements of 4-top production performed by ATLAS and CMS \cite{ATLAS:2023ajo, CMS:2023ftu} impose an upper limit on the 4-top cross section due to new physics. Comparing the cross section measured by CMS \cite{CMS:2023ftu}, $17.7^{+4.3}_{-4.0} $ fb, with the SM prediction of $13.4^{+1.0}_{-1.8}$ fb \cite{vanBeekveld:2022hty}, we find that the upper limit (at the 95\% CL) on new physics contributions to a SM-like 4-top cross section at the 13 TeV LHC is approximately 13 fb. As the cascade decays (\ref{eq:Wa}) lead to the same particles in the final state as 4-top production, we estimate a lower limit on the $t'$ mass, $m_{t'} > 0.95$ TeV, such that  $(1/4)\sigma( pp \to t' \bar t' ) < 13$ fb at $\sqrt{s} = 13 $ TeV. 
There are some caveats related to this estimated limit, because the processes (\ref{eq:Wa}) lead to a boosted $bW$ system, which in principle makes it easier for some of the event selection criteria to be satisfied, but at the same time may have the effect that some signal events are discarded by the machine learning tools employed in the analysis of \cite{CMS:2023ftu}.

\section{Conclusions}
\label{sec:conc}

Events with a large number of top quarks can be striking evidence of new physics at the LHC. While the SM production of two $t\bar{t}$ pairs has recently been measured by ATLAS \cite{ATLAS:2023ajo} and CMS \cite{CMS:2023ftu}, observing events with five or more tops at the LHC would require physics beyond the SM.
In this work, we have found that a simple and well-motivated extension of the SM by a weak-singlet vectorlike quark with electric charge $+2/3$ and a complex singlet scalar can lead to events with six tops (three $t\bar{t}$ pairs), eight tops (four $t\bar{t}$ pairs) and even ten tops (five $t\bar{t}$ pairs) at the LHC.

When the complex scalar acquires a VEV, it gives rise to two physical particles with different masses: a real scalar $\varphi_t$ and a pseudoscalar $a_t$. Mixing of the vectorlike quark with the SM up-type quark of the third generation  yields a heavy $t'$ quark, which can decay to $t a_t$ or $t \varphi_t$. 
We found that all experimental constraints on the mixing are satisfied for $s_{\! _L} \lesssim 0.1$ (see Figure~\ref{fig:sL}). Furthermore, the $t' \to t a_t/ t\varphi_t$ decays may dominate over the mixing-induced standard decay modes $t' \to th / tZ / bW$ when the mixing is relatively small, $s_{\! _L} \lesssim 0.05$, as follows from Eq.~(\ref{eq:Rt-special}).  
If the new spin-0 particles subsequently decay as $a_t \to t\bar{t}$ and $\varphi_t \to a_t a_t \to t\bar{t} t \bar{t}$, their production from $t'\bar{t}'$ pairs results in spectacular final states with many top quarks.

Specifically, the production of a $t'\bar{t}'$ pair can lead to 6-top events via the decay $t'\bar{t'} \to t\bar{t}a_ta_t$, to 8-top events via $t'\bar{t}'\to t\bar{t}a_t\varphi_t$, and to 10-top events via $t'\bar{t}' \to \varphi_t\varphi_t$. The LHC cross section for the 6-top process at $\sqrt{s}=13.6$~TeV can be up to $\sim \!20$~fb in currently unconstrained parameter space (for $m_{t'}$ near 1.1 TeV). Nevertheless, we find that the 8-top process dominates over the 6-top process for sufficiently heavy $t'$. For benchmark masses of 400~GeV and 1~TeV~for $a_t$ and $\varphi_t$, respectively, the 8-top production cross section can reach up to 2~fb (corresponding to $\sim \! 600$ events in Run 3) and surpasses the 6-top cross section for $m_{t'} \gtrsim 1.8$~TeV (see Figure~\ref{fig:xsecxbr}).

The many leptons produced in a large fraction of these events provide promising experimental handles for distinguishing signals from backgrounds. We have investigated the fraction of events from each signal process giving rise to relevant multi-lepton signatures (Section~\ref{sec:multi-lepton}), and have found that the 8-top process typically furnishes the dominant contribution to events with three same-sign leptons or five or more leptons. This underscores the potential of the 8-top process as a possible discovery mode. The cross sections for 6-top, 8-top and 10-top processes only depend on the masses of $t'$, $\varphi_t$ and $a_t$, and are independent of any couplings beyond the SM (provided that the standard $t'$ decays have negligible branching fractions). In case of a discovery, this fact allows for the measurement of the new particle masses.

Beyond the processes studied in detail in this paper, the model may give rise to various other collider signals. If the mass of $a_t$ lies below the top threshold, it preferentially decays to two gluons via a $t'$ loop. For a light $a_t$ the two-gluon system is boosted, so that the events contain a $t\bar{t}$ pair and two or more wide jets. Another interesting signal of $t'\bar{t}'$ production consists of hybrid events (see Section~\ref{sec:other}) in which either the $t'$ or the $\bar{t}'$ decays via the new modes studied in this work while the other particle decays via the standard vectorlike quark modes. This results in final states containing a $W$, $Z$ or Higgs boson in association with three or more top quarks.

 \bigskip\bigskip \bigskip   

\noindent{\it Acknowledgments: } 
We would like to thank Sekhar Chivukula and Cristian Pe\~na for helpful discussions. 
EB is supported in part by the US National Science Foundation under Grant PHY-2210177. 
Fermilab is administered by Fermi Forward Discovery Group, LLC under Contract No. 89243024CSC000002 with the U.S. Department of Energy, Office of Science, Office of High Energy Physics.

\bigskip   
  
\providecommand{\href}[2]{#2}\begingroup\raggedright

\vfil
\end{document}